\documentclass[preprint,nofootinbib, prd]{revtex4-1}
\usepackage{amsmath,amssymb,amsfonts,pgf,pgfarrows,pgfnodes,appendix,url}

\usepackage{hyperref}

\begin{document}

\title{Approximate Particle Spectra in the Pyramid Scheme}
\author{Tom Banks}
\affiliation{Department of Physics and SCIPP\\
University of California, Santa Cruz, CA 95064}
\affiliation{Department of Physics and NHETC\\
Rutgers University, Piscataway, NJ 08854}
\email{banks@scipp.ucsc.edu}
\author{T.J. Torres}
\affiliation{Department of Physics \\
University of California, Santa Cruz, CA 95064}
\email{tjtorres@ucsc.edu}

\begin{abstract}
We construct a minimal model within the general class of Pyramid Schemes\cite{pyr-scheme}, which is consistent with both supersymmetry breaking and electroweak symmetry breaking. In order to do computations, we make unjustified approximations to the low energy K\"{a}hler potential. The phenomenological viability of the resultant mass spectrum is then examined and compared with current collider limits. We show that, for certain regimes of parameters, the Pyramid Scheme can accommodate the current collider mass constraints on physics beyond the standard model with a tree-level light Higgs mass near 125 GeV. However, in this regime the model exhibits a little hierarchy problem, and one must permit fine-tunings that are generically 5\%. 
\end{abstract}

\maketitle

\section{Introduction}
	Previous work on the phenomenological implications of cosmological SUSY breaking\cite{pyr-tunnel,pyr-landau,pyr-DM} has led to the conclusion that the only class of models consistent with cosmological SUSY breaking, coupling unification and experimental bounds on gaugino masses, are the Pyramid Schemes\cite{pyr-scheme}. Though several papers\cite{pyr-tunnel,pyr-DM} have reviewed possible phenomenological effects of these models for both cosmology and high energy physics, the strong coupling gauge theory at the apex of the Pyramid made it difficult to give sharp predictions for sparticle spectra. In this paper we make several approximations, which allow us to perform rough mass spectrum calculations in a combined loop expansion/chiral effective field theory, so that we may compare them to collider limits. Although these approximations are not valid in the regime of parameters we expect to correspond to realistic models, they at least give us an indication of the sparticle spectrum in these models.
	
	We start by briefly reviewing the Pyramid Scheme as an extension to Trinification\cite{trinification-glashow,trinification}. We next outline the construction of the effective super potential as well as one-loop corrections to the K\"{a}hler metric. Finally, we show resultant mass spectra, calculated by finding the global vacuum of the scalar potential, and compare these spectra to approximate collider limits. 
 
\begin{figure}[h]
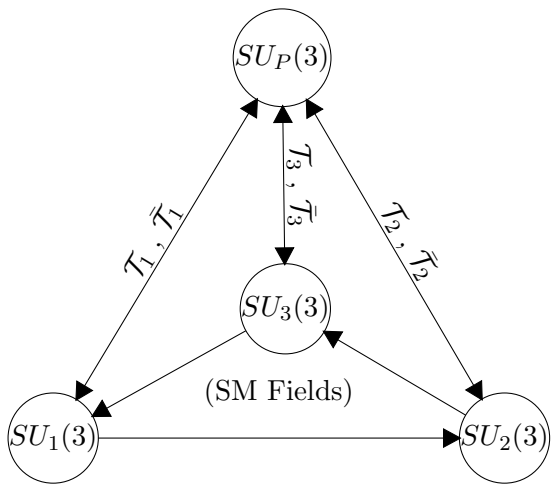

\caption{The quiver diagram of the pyramid scheme has a pyramidal shape with the base of the pyramid containing SM fields which arise from trinification, and the top of the pyramid arising from the extension of the gauge group to include $SU_P(3)$.  }
\centering
\begin{pgfpicture}{0cm}{0cm}{6cm}{6cm}
\pgfsetxvec{\pgfpoint{6cm}{0cm}}
\pgfsetyvec{\pgfpoint{.1cm}{2cm}}
\pgfsetzvec{\pgfpoint{0cm}{6cm}}

\pgfnodecircle{SU3_1}[stroke]{\pgfxyz(0,0,0)}{.6cm}
\pgfnodecircle{SU3_2}[stroke]{\pgfxyz(1,0,0)}{.6cm}
\pgfnodecircle{SU3_3}[stroke]{\pgfxyz(.5,.86,0)}{.6cm}
\pgfnodecircle{SU3_P}[stroke]{\pgfxyz(.5,.43,.7)}{.65cm}
\pgfsetendarrow{\pgfarrowtriangle{6pt}}
\pgfnodeconnline{SU3_1}{SU3_2}
\pgfnodeconnline{SU3_3}{SU3_1}
\pgfnodeconnline{SU3_2}{SU3_3}
\pgfsetstartarrow{\pgfarrowtriangle{6pt}}
\pgfnodeconnline{SU3_1}{SU3_P}
\pgfnodeconnline{SU3_2}{SU3_P}
\pgfnodeconnline{SU3_3}{SU3_P}
\pgfnodebox{SU3_1_text}[virtual]{\pgfxyz(0,0,0)}{$SU_1(3)$}{1pt}{1pt}
\pgfnodebox{SU3_2_text}[virtual]{\pgfxyz(1,0,0)}{$SU_2(3)$}{1pt}{1pt}
\pgfnodebox{SU3_3_text}[virtual]{\pgfxyz(.5,.86,0)}{$SU_3(3)$}{1pt}{1pt}
\pgfnodebox{SU3_P_text}[virtual]{\pgfxyz(.5,.43,.7)}{$SU_P(3)$}{1pt}{1pt}
\pgfnodelabelrotated{SU3_1}{SU3_P}[.5][3pt]{\pgfbox[center,base]{$\mathcal{T}_1$ , $ \bar{\mathcal{T}}_1$}}
\pgfnodelabelrotated{SU3_P}{SU3_2}[.5][3pt]{\pgfbox[center,base]{$\mathcal{T}_2 $ , $\bar{\mathcal{T}}_2$}}
\pgfnodelabelrotated{SU3_P}{SU3_3}[.5][3pt]{\pgfbox[center,base]{$\mathcal{T}_3 $ , $\bar{\mathcal{T}}_3$}}
\pgfputat{\pgfxyz(.5,0,.1)}{\pgfbox[center,center]{(SM Fields)}}

\end{pgfpicture}
\label{Pyramid}
\end{figure}

	The pyramid scheme is an effective theory that extends Trinification to include a fourth, strongly coupled, $SU(3)$ factor. Thus the full gauge group of the theory is given by $\mathcal{G} = SU_1(3) \times SU_2(3) \times SU_3(3) \times SU_P(3) \ltimes \mathbb{Z}_3$, wherein we associate the $SU_3(3)$ factor with standard model color, $SU_2(3) \times SU_1(3) \longrightarrow SU_L(2) \times U_Y(1)$ after GUT spontaneous symmetry breaking, and $SU_P(3)$ is the aforementioned fourth gauge factor, which we call the pyramid $SU(3)$. In addition, the theory is complete in its description of symmetry breaking in the sense that all relevant soft SUSY breaking terms are accounted for via explicit interactions in the theory, whether through F-term contributions to the scalar potential or gauge mediated loop corrections.  We remind the reader that the Pyramid Scheme is {\it unnatural} from the point of view of effective field theory.  Interactions which break a discrete R-symmetry are imagined to have originated from a special class of diagrams in which a gravitino propagates from the vicinity of a particular local observer to the horizon, interacts with the non-field theoretic degrees of freedom there and returns to the vicinity of that observer.   We will discuss the discrete R-symmetries of our particular model in the appendix.   In addition to the gauge and matter content summarized in the quiver diagram of Fig. \ref{Pyramid}, the model contains gauge singlets $S_i$, which are essential for implementing SUSY breaking.  The minimal number is $3$ and we will work with that minimal content in this paper.

	The origin of the name Pyramid Scheme is evident in the quiver diagram of Fig. \ref{Pyramid}, where standard model generations run around the base of the pyramid and additional field content is given by:
\begin{center}
\begin{math} 
\begin{array}{c|cccc}
 & SU_1(3) & SU_2(3) & SU_3(3) & SU_P(3)\\\hline
 S_i & 1 & 1 & 1 & 1\\
{\cal T}_1 & \Box & 1 & 1 & \bar{\Box}\\
\bar{\cal T}_1 & \bar{\Box} & 1 & 1 & \Box\\
{\cal T}_2 & 1 & \Box & 1 & \bar{\Box}\\
\bar{\cal T}_2 & 1 & \bar{\Box} & 1 & \Box \\
{\cal T}_3 & 1 & 1 & \Box & \bar{\Box}\\
\bar{\cal T}_3 & 1 & 1 & \bar{\Box} & \Box
\end{array}
\end{math}
\end{center}

Here the ${\cal T}_i$ are fields which transform in the bifundamental of $SU_i(3) \times SU_P(3)$ which we call "trianons," and the $\bar{\cal T}_i$ are the conjugates of the ${\cal T}_i$.

	Though $SU_P(3)$ must be strongly coupled at the TeV scale, it is not asymptotically free at high energies. $SU_P(3)$ does become asymptotically free below the highest trianon mass scale, and thus we seek to look at effective field theories below this scale such that at low energies all field content is $SU_P(3)$ confined. Additionally, we set the confinement scale, $\Lambda_P$, to be below the second heaviest trianon so as to simplify calculations by allowing the use of Seiberg's $N_F=N_C$ formalism \cite{SeibergD} below the confinement scale, when the two heavier trianons are integrated out.  The trianon masses violate R-symmetry, and are expected to be at the terascale.
	
If any of the trianons are much heavier than this, the corresponding standard model gaugino will be very light, and this is a contradiction with experimental bounds.  The problem of getting $\Lambda_P$ close to the heaviest trianon mass was studied in \cite{pyr-scheme}.  If one wants to preserve an approximate pyrma-baryon number, it requires $\Lambda_P$ to be close to, perhaps a little less than, the TeV scale.

As noted above, the fact that the trianons are charged under the standard model, allows for gauge mediation in the effective theory. It also provides a source for non-canonical contributions to the K\"{a}hler potential. Using these simplifications, we can build a minimal, gauge-invariant, effective theory with which to calculate the Chargino, Gaugino, pyramid fermion, and pyramid boson masses of the pyramid scheme as well as first order corrections to the K\"{a}hler metric. 

\section{Model Building}
Above the pyramid confinement scale the minimal, gauge-invariant superpotential that breaks supersymmetry is
\begin{equation}
W= \alpha^i S_i + (\beta^i S_i + \mu)H_u H_d + (\gamma^{i k} S_i + m^{ k }) \mathcal{T}_k \tilde{\mathcal{T}}_k + \sum_k  g^{k}\left[ (T_{k})^3 + ( \tilde{\mathcal{T}}_{k} )^3 \right]+W_{SM} \nonumber
\end{equation}
with $(T_{k})^3 = \epsilon_{a_1 a_2 a_3} \epsilon^{n_1 n_2 n_3} (\mathcal{T}_k)^{a_1}{}_{n_1} (\mathcal{T}_k)^{a_2}{}_{n_2} (\mathcal{T}_k)^{a_3}{}_{n_3} $ being the gauge-invariant cubic combination of the trianons and $W_{SM}$ representing the standard model contribution to the superpotential. The superpotential contains a $\mu$-term for the Higgs doublets as well as singlets that couple to the Higgs. The theory has many of the same features as the NMSSM. Moreover, the singlets have linear terms in the superpotential so as to facilitate O'Raifeartaigh breaking in the singlet sector.  The SUSY breaking does not decouple from the standard model due to the singlet couplings to the Higgs doublets, and the standard model gauge couplings of the trianons. 

	Examining the effective theory below $\Lambda_P$, we identify the gauge-invariant polynomials that will be fundamental below the scale of $SU_P(3)$ confinement
\begin{eqnarray}
\operatorname{Tr} \mathbf{M}_{k} & = & (\mathcal{T}_{k }){}^{a}{}_{n} (\tilde{\mathcal{T}}_{k}){}_{ a}{}^{n}  \nonumber \\
B_k &=& \epsilon_{a_1 a_2 a_3} \epsilon^{n_1 n_2 n_3} (\mathcal{T}_k)^{a_1}{}_{n_1} (\mathcal{T}_k)^{a_2}{}_{n_2} (\mathcal{T}_k)^{a_3}{}_{n_3} \nonumber \\
\tilde{B}_k &=& \epsilon^{a_1 a_2 a_3} \epsilon_{n_1 n_2 n_3} (\tilde{\mathcal{T}}_k)_{a_1}{}^{n_1} (\tilde{\mathcal{T}}_k)_{a_2}{}^{n_2} (\tilde{\mathcal{T}}_k)_{a_3}{}^{n_3} \nonumber
\end{eqnarray}
We choose the first two trianons, $\mathcal{T}_{1,2}$,  to be above $\Lambda_P$. They can  be integrated out of the effective theory, leaving only the trianon coupled to standard model color.  We do this in order to ensure that the effective  theory below $\Lambda_P$ has a simple effective Lagrangian description.  If we had two light trianons,  the theory would be in the superconformal window and difficult to analyze.  

We choose the single light trianon to be that associated with color in order to account for the experimental bounds on the gluino mass.   Standard model gauginos get their masses from gauge mediation, with trianons as messengers.  If the supersymmetric mass of a given trianon is taken much heavier than $\Lambda_P$ then the corresponding gaugino mass goes to zero.  Since the strongest experimental bounds are those for gluinos, we insist that the the colored trianon be the lightest.  Note that the heavier trianon masses cannot be taken arbitrarily large, because this would make charginos very light.  We imagine that they are probably of order a few times $\Lambda_P$, for example something like the mass of the $\rho$ meson in QCD confinement scale units.  The mass of the colored trianon is more like that of the strange quark, small enough for Seiberg's chiral effective Lagrangian to be a good approximation.  Below, when we calculate corrections to the K\"{a}hler potential from integrating out the heavier trianons, we will neglect the $SU_P (3)$ gauge interactions.  This would be a good approximation for trianon masses $>> \Lambda_P$, which would not give an acceptable chargino spectrum.  We cannot do reliable calculations in the regime which is phenomenologically relevant, so the detailed numerical results of the present paper should be taken only as indications of the properties of the Pyramid Scheme.

Knowledgeable readers may worry that there is not enough of a logarithmic distance between the heavy trianon masses and $\Lambda_P$.  The latter scale is generated by the renormalization group once the running of the gauge coupling stops being IR free.  This problem was addressed in \cite{pyr-landau}.  The model has an attractive fixed line where all three couplings $g^k$ are non-zero and equal to the gauge coupling of $SU_P (3)$.  If we were on this fixed line close to the end of the perturbative regime, $\Lambda_P$ would be very close to the heavy trianon masses. If only two of the couplings are non-zero there is no more fixed line, but the running of the gauge coupling is slow enough to be consistent with a small ratio between the masses and $\Lambda_P$, if $\Lambda_P$ is about $900$ GeV or less.  We would like at least one of the $g^k$ to be zero in order that we have an approximate baryon number symmetry in the trianon sector, which gives the model a dark matter candidate.  To simplify the low energy analysis, we have chosen $g^3 \neq 0$.   As a consequence, the dark matter candidate is made out of ${\cal T}_i$ with $i = 1$ or $2$, and it has a magnetic dipole moment.  Its mass is probably between $ 5 - 20 $ TeV.  

From these considerations, we conclude that the effective theory below $\Lambda_P$ contains only one pyramid baryon, and its conjugate, as well as the trace of the meson matrix:
 
\begin{eqnarray}
\operatorname{Tr} \mathbf{M} & = & (\mathcal{T}_{3 }){}^{a}{}_{n} (\tilde{\mathcal{T}}_{3}){}_{ a}{}^{n}  \nonumber \\
B &=& \epsilon_{a_1 a_2 a_3} \epsilon^{n_1 n_2 n_3} (\mathcal{T}_3)^{a_1}{}_{n_1} (\mathcal{T}_3)^{a_2}{}_{n_2} (\mathcal{T}_3)^{a_3}{}_{n_3} \nonumber \\
\tilde{B} &=& \epsilon^{a_1 a_2 a_3} \epsilon_{n_1 n_2 n_3} (\tilde{\mathcal{T}}_3)_{a_1}{}^{n_1} (\tilde{\mathcal{T}}_3)_{a_2}{}^{n_2} (\tilde{\mathcal{T}}_3)_{a_3}{}^{n_3} \nonumber
\end{eqnarray}

	We neglect the standard model contribution to the effective superpotential and institute Seiberg's quantum moduli constraint for $N_F=N_C$ to arrive at
\begin{equation}
W_{eff} = \alpha^i S_i + (\beta^i S_i + \mu) H_u H_d + (\gamma^i S_i + m)\operatorname{Tr} \mathbf{M} + g_{P_1}B + g_{P_2} \tilde{B} + X(\det{\mathbf{M}} -B \tilde{B} - \Lambda_P^6)\nonumber
\end{equation}
where $X$ is a new Lagrange multiplier field included to enforce the constraint, $\det{\mathbf{M}} -B \tilde{B} = \Lambda_P^6$.
 
	Furthermore, extensive numerical tests have shown that the vacuum of the effective theory preserves color. Thus, we can make the additional simplification that $\mathbf{M}$ be diagonal and given by
\begin{equation}
\mathbf{M}^i{}_j = 
\begin{pmatrix}
M & 0 & 0\\
0 & M & 0 \\
0 & 0 & M
\end{pmatrix} \nonumber
\end{equation}
Inserting this form of the meson matrix into the effective potential gives the final form
\begin{equation}
W_{eff} = \alpha^i S_i + (\beta^i S_i + \mu) H_u H_d + 3 (\gamma^i S_i + m) M + g_{P_1}B + g_{P_2} \tilde{B} + X(M^3 - B \tilde{B} - \Lambda_P^6)\nonumber
\end{equation}

	SUSY breaking appears in the singlet sector via non-zero F-term contributions. This can be seen by examining the F-terms:
\begin{eqnarray}
F_{S_i} & = & \alpha^i + \beta^i H_u H_d + 3 \gamma^i M \nonumber \\
F_{H_u} & = & (\beta^i S_i + \mu) H_d \nonumber \\
F_{H_d} & =& (\beta^i S_i + \mu) H_u \nonumber \\
F_M & = & 3 (\gamma^i S_i + m +  X M^2) \nonumber \\
F_B & = & g_{P1} + X \tilde{B} \nonumber \\
F_{\tilde{B}} & = & g_{P2} + X B \nonumber \\
F_X & = & M^3 - B \tilde{B} - \Lambda_P^6 \nonumber
\end{eqnarray}
For the singlet sector we can see that if we have $n$ singlets and the parameter $n$-vectors ($\alpha^i$, $\beta^i$, $\gamma^i$) are linearly independent, then we have $n$ equations in two effective variables, giving no solution when $n>2$. For our purposes, we will be interested in the minimal case when $n=3$, but additional models with $n>3$ are possible.\footnote{ It should be noted as well that, despite our simple approach, one can also add certain quadratic and cubic interaction terms for the singlets to the superpotential so long as they don't ruin the SUSY breaking structure. }

	The theory is additionally simplified by noting that the F-term equations for $B$, $\tilde{B}$, and $X$ can be set to zero without loss of generality. These fields do not appear anywhere else in the superpotential, and their F terms make a positive definite contribution to the potential. This allows us to solve for the VEVs of those fields in terms of $M$:
\begin{eqnarray}
B &=& \sqrt{ \frac{g_{P_2}(M^3 - \Lambda_P^6)}{g_{P_1}}} \nonumber \\
\tilde{B} &=& \sqrt{ \frac{g_{P_1}(M^3 - \Lambda_P^6)}{ g_{P_2}}} \nonumber \\
X &=& \sqrt{ \frac{g_{P_1} g_{P_2}}{M^3 - \Lambda_P^6}} \nonumber 
\end{eqnarray}
With this simplification the effective theory has six effective complex degrees of freedom and 14 parameters at the level of the superpotential. All that is needed to fully determine the effective scalar potential is then the inclusion of D-terms and knowledge of the K\"{a}hler potential.

\section{D-terms and the Higgs Potential}
The D-term contributions to the scalar potential arise from the Higgs sector and are given by
\begin{eqnarray} 
{1 \over 2} \sum_a D^a D^a  & = & {1 \over 2} \sum_a g_a^2 ( \phi^* T^a \phi)^2 \nonumber \\
&=& {1 \over 8} (g^2+g'^2) (|H^+_u|^2 + |H^0_u|^2 - |H^-_d|^2 - |H^0_d|^2)^2 + {1 \over 2}| H^+_u H^{0 *}_d + H^0_u H^{- *}_d |^2 .\nonumber
\end{eqnarray}
This gives the full Higgs potential of the pyramid scheme as
\begin{eqnarray}
V_{\mbox{Higgs}} &=& \left[ \left( \alpha^i + \beta^i (H_u^+ H_d^- - H_u^0 H_d^0 ) + 3 \gamma^i M \right) \beta_i^* (H_u^{+ *} H_d^{- *} - H_u^{0 *} H_d^{0 *} ) + \mbox{h.c.} \right] \nonumber \\
&& + {1 \over 8} (g^2+g'^2) (|H^+_u|^2 + |H^0_u|^2 - |H^-_d|^2 - |H^0_d|^2)^2 + {1 \over 2}| H^+_u H^{0 *}_d + H^0_u H^{- *}_d |^2 .\nonumber
\end{eqnarray}

	In order to show that the vacuum preserves electromagnetism , we can use an $SU(2)$ gauge transformation to set $H^+_u = 0$ and then  $\partial V / \partial H^+_u=0$ implies that $H^-_d = 0$. After setting $H^+_u = H^-_d = 0$ one notes that there is a symmetry under the interchange of the neutral Higgs components, $H^0_u \leftrightarrow H^0_d$. 

	This symmetry is not spontaneously broken, and requires that $\tan \beta=1$ in the vacuum, which is inconsistent with the requirement that the top-quark Yukawa coupling remain perturbative up to unification. In order to satisfy this requirement, we note that there are large radiative corrections from top/stop loops which give rise to 
\begin{equation}
\delta V_{eff} = - {12 y_t^2 \over 16 \pi^2} |H_u|^2 m_{\tilde{t}}^2 \ln(\Lambda_c / m_{\tilde{t}}) \nonumber
\end{equation}
This contribution favors large $\tan \beta$ and including it allows for $\tan \beta > 1.7$, the perturbative unification bound, as long as the stop mass is not too small.

\section{Effective K\"{a}hler Metric}
The final piece necessary to calculate the effective scalar potential is the contribution to the effective K\"{a}hler metric given by integrating out ${\cal T}_{1,2}$. In order to calculate the non-canonical pieces of the K\"{a}hler metric we first take
\begin{equation}
\int d^2 \theta \; (a^i S_i + m_1) \mathcal{T}_1 \tilde{\mathcal{T}}_1 + (b^i S_i + m_2) \mathcal{T}_2 \tilde{\mathcal{T}}_2 \nonumber
\end{equation}
and integrate out $\mathcal{T}_{1,2}$, and then match the resultant expression term-by-term with $F_{i} \bar{F}_{\bar{\jmath}} \mathcal{K}^{i \bar{\jmath}}$.

Thus we have that
\begin{eqnarray}
\int d^2 \theta \; W &=& \int d^2 \theta \; (a^i S_i + m_1) \mathcal{T}_1 \tilde{\mathcal{T}}_1 + (b^i S_i + m_2) \mathcal{T}_2 \tilde{\mathcal{T}}_2\nonumber \\ 
&=& \int {d^4 p \over (2 \pi)^4} \; \Bigg[ \ln \left( p^2 - m_1^2 - a^i \bar{a}^{\bar{\jmath}} S_i \bar{S}_{\bar{\jmath}}  \mp \sqrt{a^i \bar{a}^{\bar{\jmath}}}  F_{i} \bar{F}_{\bar{\jmath}} \right) \nonumber \\
&& -  2 \ln \left(p^2 - m_1^2 - a^i \bar{a}^{\bar{\jmath}} S_i \bar{S}_{\bar{\jmath}} \right) \nonumber \\
&& +  \ln \left( p^2 - m_2^2 - b^i \bar{b}^{\bar{\jmath}} S_i \bar{S}_{\bar{\jmath}}  \mp \sqrt{a^i \bar{a}^{\bar{\jmath}}}  F_{i} \bar{F}_{\bar{\j}} \right) \nonumber \\
&& -  2 \ln \left(p^2 - m_1^2 - b^i \bar{b}^{\bar{\jmath}} S_i \bar{S}_{\bar{\jmath}}\right) \Bigg] \nonumber 
\end{eqnarray}
where $i$ and $j$ range over the $S_i$'s. And expanding the log, we have
\begin{equation}
F_{i} \bar{F}_{\bar{\jmath}} \mathcal{K}^{i \bar{\jmath}} \approx -   F_{i} \bar{F}_{\bar{\jmath}} \int {d^4 p \over (2 \pi)^4} \left(  {a^i \bar{a}^{\bar{\jmath}} \over (p^2 - a^i \bar{a}^{\bar{\jmath}} S_i \bar{S}_{\bar{\jmath}})^2} +  {b^i \bar{b}^{\bar{\jmath}}  \over (p^2 - b^i \bar{b}^{\bar{\jmath}} S_i \bar{S}_{\bar{\jmath}})^2} \right)\nonumber
\end{equation}

	We calculate the integral using dimensional regularization in the $\overline{MS}$ scheme resulting in a K\"{a}hler metric which has the form:
\begin{equation}
\mathcal{K}^{i \bar{\jmath}} =
\delta^{i \bar{\jmath}} - \frac{1}{8\pi^2} \left[ a^i \bar{a}^{\bar{\jmath}} \ln  \left| m_1 + a_i S^i \right| + b^i \bar{b}^{\bar{\jmath}}  \ln \left| m_2 + b_i S^i \right| \right] \nonumber
\end{equation}
where $a^i = b^i=0$ for $i \in \{H_u, H_d, M, B, \tilde{B}, X\}$.  

We emphasize again that this approximation to the K\"{a}hler potential is not valid in the regime of parameters in which we actually work.

\section{Mass Spectrum Calculation}

The mass spectrum of the theory can now be computed by diagonalizing the following mass matrices:
\begin{eqnarray}
\mathcal{M}_F &=& W_{i j} = D_i D_j W = \left( \partial_i \partial_j - \Gamma^k_{i j} \partial_k \right) W\nonumber \\
\nonumber \\
 \mathcal{M}_B^2 &=&
\begin{pmatrix}
W_{jk}^* W^{ik} & W_{ijk}^* W^k \\
W^{ijk} W^*_k & W^*_{ik} W^{jk}
\end{pmatrix}  \nonumber \\
\nonumber \\
\mathcal{M}_C &=& 
\begin{pmatrix} 
 0 & 0 & Y_2 {\alpha_2 \over 4 \pi} \sum_i \left| {F_i \over \Lambda_P} \right| & g \langle H_d^{0*} \rangle \\
 0 & 0 & g \langle H_u^{0*} \rangle & (\langle \beta^i  S_i \rangle + \mu ) \\
 Y_2 {\alpha_2 \over 4 \pi} \sum_i \left| {F_i \over \Lambda_P} \right| & g \langle H_u^{0*} \rangle & 0 & 0 \\
 g  \langle H_d^{0*} \rangle & ( \langle \beta^i  S_i \rangle + \mu ) & 0 & 0 
 \end{pmatrix}  \nonumber \\
 \nonumber \\
  \mathcal{M}_g &=&  Y_3 {\alpha_s \over 4 \pi}  \sum_i \left| {F_i \over \Lambda_{P}} \right| \nonumber
 \end{eqnarray}
where $Y_2$ and $Y_3$ are numerical constants of order one (which we set to one for purposes of calculation), the $F_i$ are the F-term values in the vacuum, and $\Gamma^k_{i j} = \mathcal{K}^{k \bar{\imath}}  \mathcal{K}_{i \,\bar{\imath} ,\, j} $ is the K\"{a}hler connection. 

We have included gauge mediated Majorana gaugino masses in these formulae, with unknown order one coefficients.  The loop diagrams which generate these terms involve strongly interacting particles.

	As the theory contains a large number of complex fields, analytic diagonalization is computationally prohibitive. Thus we must first find the vacuum and then numerically diagonalize the mass matrices to obtain the mass spectrum. This has the advantage of being fast, but the disadvantage of obscuring the structure of the mass eigenstates. 

	In order to minimize the scalar potential we first choose the confinement scale, $\Lambda_P$, to be one and pick values for the coupling parameters. All experimentally known couplings are taken to be equal to their values at the weak scale, and RG running is not taken into account in the computation. We then seek a global minimum by employing a random search algorithm\footnote{Although a random search is not very elegant, we performed many tests over a vast range of different algorithms (including differential evolution, simulated annealing, firefly, etc.) and within these tests a random search consistently found lower minima in less time in comparison to other algorithms. }. After finding the minimum, $\Lambda_P$ is then set to the value which gives the correct Z mass via
\begin{equation}
\Lambda_P = {174 \over \sqrt{ v_u^2 + v_d^2 }} \mbox{GeV} \nonumber
\end{equation}
Where here $v_u$ and $v_d$ are the dimensionless numbers corresponding to the Higgs VEVs. We then take our dimensionless numerical data for masses, and all massive quantities are rescaled by appropriate powers of $\Lambda_P$, using the above formula to get answers in GeV (\emph{i.e.} $v_{u,d} \; \Lambda_P = \left| \langle H_{u,d} \rangle \right|$).

\section{Results}

\begin{figure*}
\center
\includegraphics[width= .8 \textwidth]{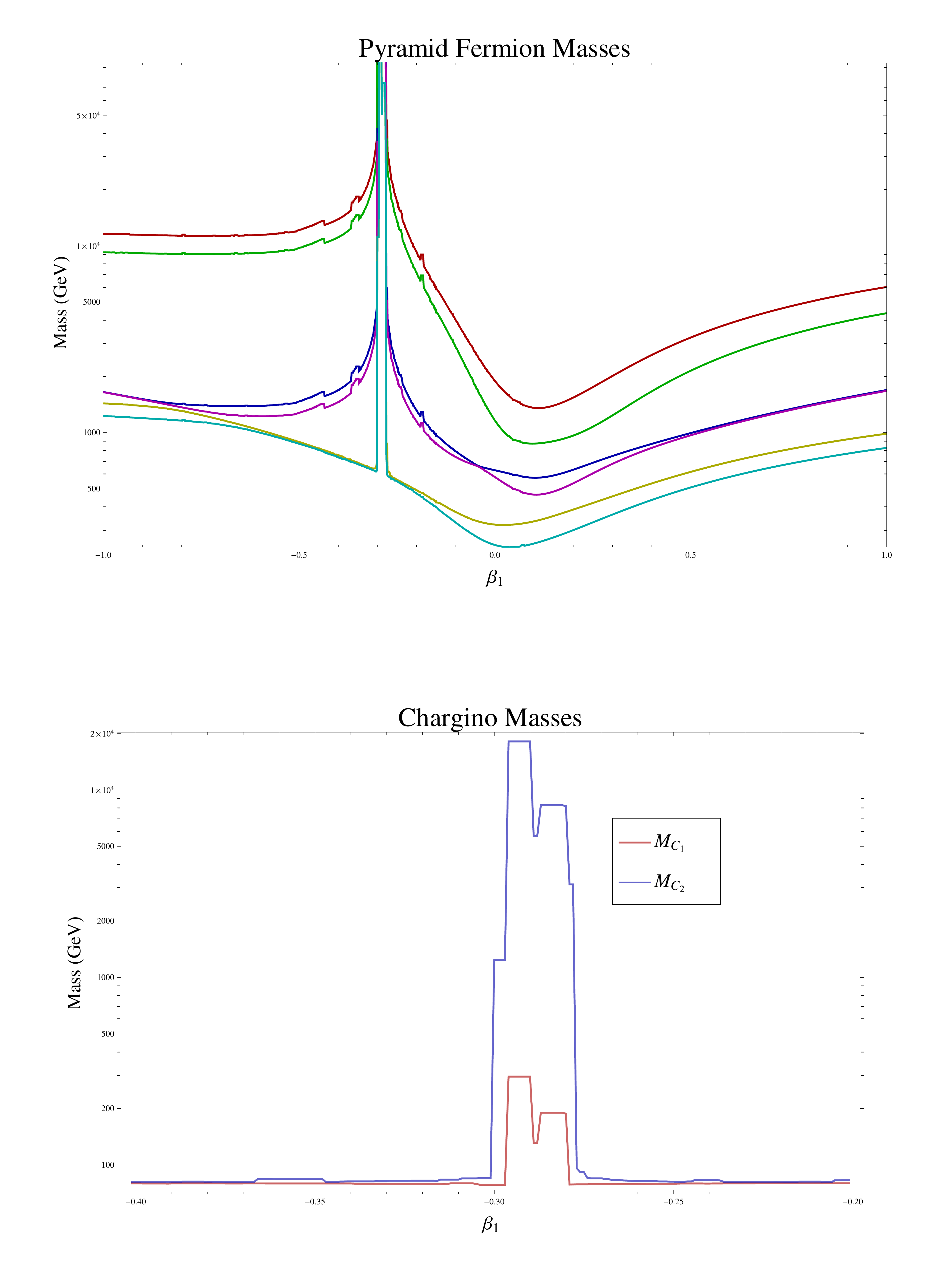}
\caption{Fermion and chargino masses are shown as a function of the coupling parameter, $\beta_1$. The valid mass regime can be seen in the region of the large mass spikes giving chargino masses just above the experimental bound near 300 GeV. Separate parameters provide even higher mass regimes for the light chargino.}  
\label{masses1}
\end{figure*}

\begin{figure*}
\center
\includegraphics[width= \textwidth]{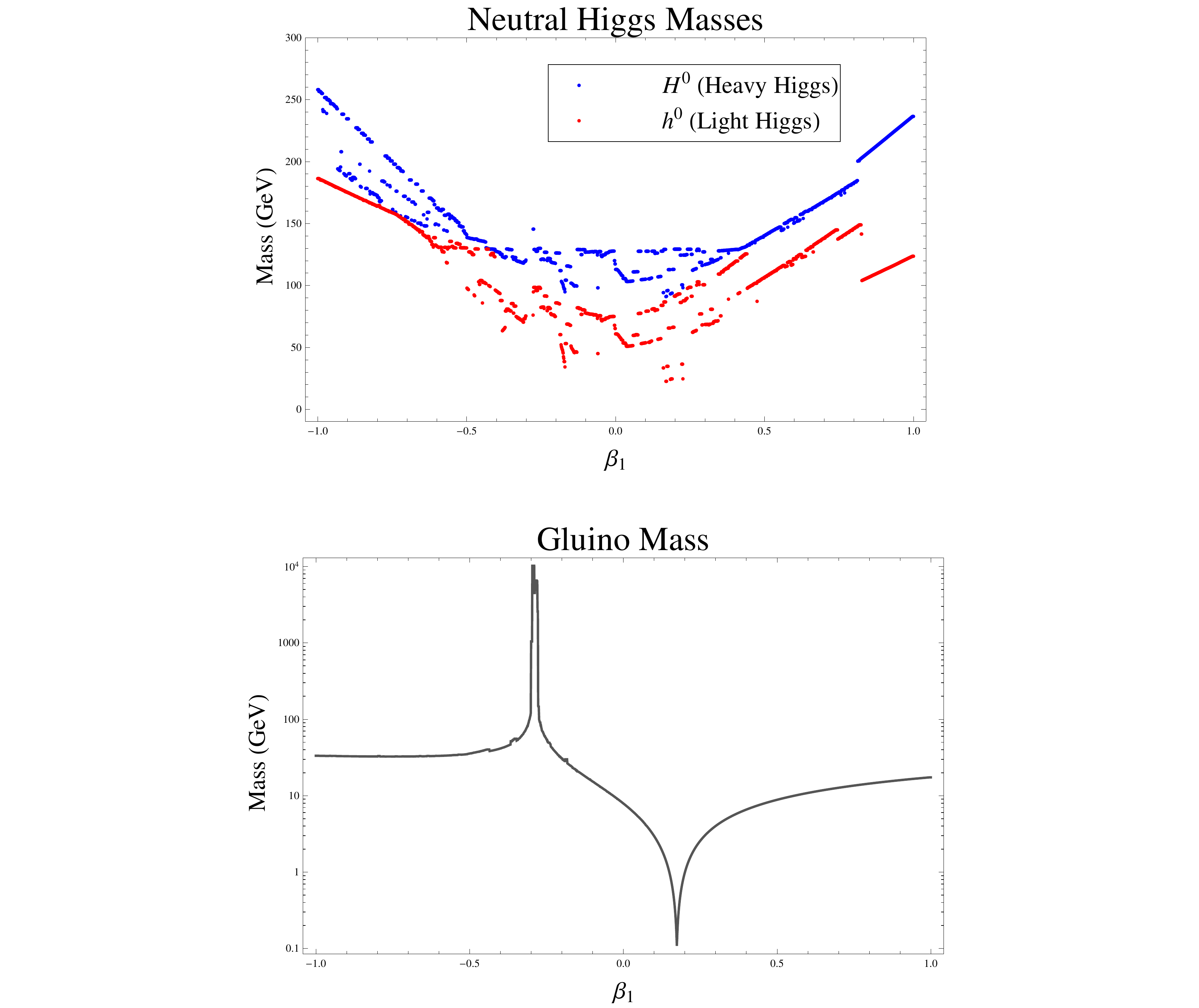}
\caption{ Neutral Higgs and gluino masses are shown as functions of $\beta_1$.  The tree-level light Higgs has a mass range of roughly 60-180 GeV over the $\beta_1$ scan. Additionally, the valid mass regime of the gluino can be seen in the region of the large mass spike. This represents a moderate degree of fine-tuning in order to set the correct mass hierarchy. }
\label{masses2}
\end{figure*}

In comparing the mass spectrum of the theory to experimental limits we use the the following rough cutoffs as mass bounds:
\begin{equation}
\begin{centering}
\begin{array}{cc}
\mbox{Pyramid Fermions} & > 300\, \mbox{GeV}\\
\mbox{Neutral Higgs} &  125 \,\mbox{GeV} > m_h > 100 \, \mbox{GeV} \, , \, m_H> 300 \, \mbox{GeV} \\
\mbox{Charginos} & > 300 \, \mbox{GeV}\\
\mbox{Gluino} & > 600 \, \mbox{GeV}
\end{array} \nonumber
\end{centering}
\end{equation}

	The results from the mass calculation are summarized in Figures \ref{masses1} and \ref{masses2}. One can see that there are valid regions of parameter space for which all experimental mass bounds are satisfied. However, one may also note that such bounds are only satisfied over a region of parameter space where the masses change quickly as a function of the couplings. This represents a little hierarchy problem, which requires a sizable degree of fine-tuning. 
	
	Note that we have provided graphs only for the variation of a few masses with the single parameter $\beta_1$.  We've actually generated complete spectra for a reasonable range of {\it all} of the parameters of the model, and used the conventional fine tuning measure, which emphasizes the parameter to which any given mass is most sensitive.  It is not always $\beta_1$.  We've displayed the gluino prominently, since it is the particle with the most stringent experimental lower bounds and therefore gives the most tension with the correct $Z$ and $W$ masses.  Our data and the simple software tools needed to analyze them may be found on the following website in \cite{web1}.

\section{Little Hierarchy Problem}
The little hierarchy in the model stems from the difference in scales necessary between the pyramid and weak scales in order to simultaneously satisfy the experimental mass bounds on the yet unobserved particles and the condition that the Higgs VEVs be such that they give correct values for the W and Z masses. In effect, one can see that this will be the case in the mechanics of the calculation, wherein we fix $\Lambda_P$ to give the correct values of the W and Z masses. Then, since other masses are scaled by $\Lambda_P$ this sets the requirement that $ \langle H^0_u \rangle$ and  $ \langle H^0_d \rangle$ be rather small in units of the pyramid scale, thus requiring a hierarchy. 

\begin{figure*}
\centering
\includegraphics[width=5.0in]{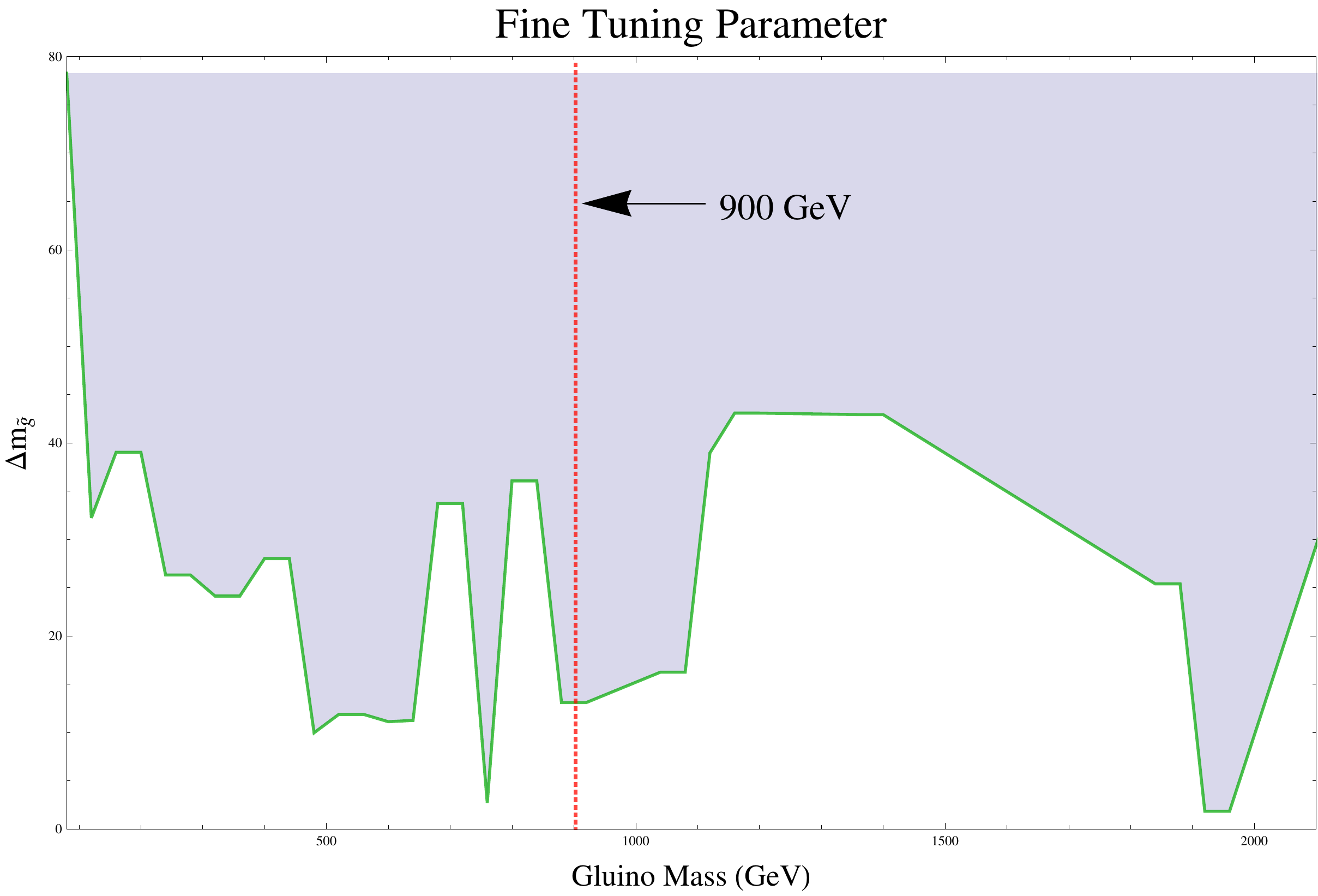} 
\caption{Plot of the fine-tuning parameter of the gluino mass ($\Delta_{m_{\tilde{g}}}$) as a function of gluino mass ($m_{\tilde{g}}$). This plot was generated by calculating fine-tuning parameters for all parameters over a single cross section of parameter space, collecting the results into mass bins of gluino mass, and then finding the maximum fine-tuning over all parameters within each mass bin. Above the mass bound at 900 GeV the fine-tuning parameter varies between 20 and 40 corresponding to a fine-tuning of roughly 2.5-5\%. The maximum fine-tuning occurs at low mass and is marked by the blue border near the top of the plot.  }
\label{fine-tuning}
\end{figure*}

	It will be advantageous to calculate the extent of the hierarchy by computing the fine-tuning necessary to satisfy the above experimental constraints. To do this we use the (standard) metric \cite{fine-tuning}
\begin{eqnarray}
\Delta_{m_a} &=& \max_i \left| { \partial \ln m_{\tilde{g}} \over \partial \ln \lambda_i} \right| 
\nonumber \\
\lambda_i &=& \{a_i, b_i,  \alpha_i, \beta_i, \gamma_i, \mu, g_{P1}, g_{P2}, m_i \} \nonumber
\end{eqnarray}
The degree of fine-tuning to give any particular mass higher than a certain bound will be $1/ \Delta_{m_a}$ evaluated at the points which give the mass in question, $m_a$,  a value greater than the lower bound of the experimental constraints. In the mass spectrum above, the greatest level of fine-tuning stems from requiring that the gluino and lightest chargino masses be above current collider bounds. We have determined through many trials that the gluino mass generally governs the maximum fine-tuning in these models and a plot of $\Delta_{m_{\tilde{g}}}$ as a function of $m_{\tilde{g}}$ can be found in Fig. \ref{fine-tuning}.

We find that the fine-tuning of minimal models in the pyramid is roughly of order 2.5-5\%, corresponding to a $\Delta m_{\tilde{g}} \approx 20-40$.

\section{Conclusions}

We have shown that, granted our uncontrolled approximations, the pyramid scheme admits a mass spectrum of yet undiscovered particles which is allowed by current collider limits. Though the approximations made in the calculation of the K\"{a}hler metric were without justification, they have allowed us to calculate mass spectra which should be indicative of minimal models in the pyramid scheme without the inclusion of soft SUSY-breaking terms of unknown origin. The numerical results presented are approximate, but show promise for the viability of cosmological SUSY breaking as a theory of nature. The fine tuning is not so severe.  We also remind readers that many things in nuclear physics appear highly tuned.  That tuning is explained, in a complicated way, by the strong coupling physics of QCD.  The Pyramid Scheme has similarly complicated strong coupling physics, and we might in the end have simply to accept this until such time as efficient lattice calculations in supersymmetric gauge theories can test the real predictions of the model.  Still, it would be interesting to find a simple mechanism in the model that avoids tuning altogether.

In the MSSM, fine-tuning stems from the tension between achieving a high (125 GeV) Higgs mass and the correct scale of electroweak symmetry breaking. For example, Hall et. al. \cite{Halletal} have studied the necessary fine-tunings in the MSSM to bring the Higgs mass up to the LHC excess region, with results that indicate fine-tuning of at least $\sim 1\%$, even with maximal mixing. In the pyramid scheme, the case is different with a natural tree-level light Higgs mass in the range of 60-180 GeV. Essentially, this is because the Pyramid scheme incorporates singlets with NMSSM-like couplings. 

Nonetheless, fine-tuning still arises in the Pyramid Scheme from the tension between having the correct electroweak breaking scale and having chargino/gluino masses that lie above the LHC exclusion bounds. 

We note that several issues of RG running are modified, in comparison with the MSSM, in the Pyramid Scheme.  With the effective potential used in the current paper, we find $\tan\beta = 1$ , which would make the top Yukawa have a Landau pole below the unification scale.  We note that most NMSSM models prefer fairly small values of $\tan\beta$.  The contribution to the effective potential from top-stop splitting will be enough to drive $\tan\beta$ above its perturbative unification bound if the stop mass is in the TeV range.  This is the natural range of values for $m_{\tilde{t}}$ in the Pyramid scheme.

Furthermore, while {\it e.g.} \cite{Halletal} find that naturalness in the NMSSM prefers a value for the singlet coupling to the Higgs that drives this coupling non-perturbative below the unification scale, the Pyramid scheme has three different singlets, each with its own coupling to the Higgs.  It seems plausible that this will allow the couplings to remain perturbative, even for a $125$ GeV Higgs.  Of course, the current version of our model is slightly ``unnatural" by the criteria of \cite{Halletal}.

\subsection{Review and summary}

To conclude, we review the basic features of the Pyramid Schemes, their successes, and the place of the current paper in the study of these models.  The models are motivated by the Holographic Space-time (HST) theory of stable dS space.  According to that theory, there is a model for every sufficiently small value of the cosmological constant.  As $\Lambda \rightarrow 0$, SUSY is restored, along with a discrete R symmetry, according to the formula
$$m_{3/2} = K \sqrt{m_P / m_U} \Lambda^{1/4},$$ with $m_P$ the reduced Planck mass, $m_U$ the unification scale and $K$ a constant of order one.  $m_U$ is usually taken to be $ 2\times 10^{16}$ GeV.

The Pyramid schemes are low energy field theories (LEFTs) , which are constructed to implement the results of HST.  The very low SUSY breaking scale implied by the formula for the gravitino mass, along with coupling unification and experimental bounds on sparticle masses, puts strong constraints on these theories.  In addition, theoretical constraints \cite{pyr-tunnel}, imply that these models cannot satisfy the constraints of field theoretic naturalness. The $\Lambda = 0$ LEFT preserves SUSY and a discrete R symmetry.  The corrections to it come from interactions of a single gravitino with degrees of freedom on the dS horizon.  They break the discrete R symmetry and must lead to spontaneous symmetry breaking.  The Nelson-Seiberg theorem\cite{nelsonseiberg} implies that these terms cannot be generic, but their very special origin in graphs where a single gravitino propagates to the horizon and returns, suggests that they are indeed non-generic.  Corrections to the single gravitino diagrams are {\it exponentially} suppressed. This has the following ``unnatural" consequences

\begin{itemize}

\item The discrete R symmetry sets the constant in the superpotential to zero, so that Poincare invariance is restored along with SUSY.  The R violating constant is tuned, in order to implement
the above gravitino mass relation.

\item The discrete R symmetry is chosen to eliminate all B and L violating operators of dimension $4$ and $5$ from the LEFT.  One can argue that the gravitino exchange diagrams are UV insensitive and will reintroduce these terms only with powers of the TeV scale over the unification or Planck scale.  The results are compatible with experimental bounds. The Pyramid Schemes are among a small class of extant models, which deal with generic B and L violation through dimension $5$.

\item The $\Lambda = 0$ LEFT has $\theta_{QCD} = 0$ as a consequence of an anomalous $U(1)$.  The R violating terms lift the would be axion to a TeV scale mass, but, as a consequence of their special origin, do not introduce new CP violating phases, beyond those induced by standard model loops via the CKM matrix.  This is a novel solution of the strong CP problem.

\item The R symmetry can be chosen to forbid the conventional $\mu$ term.  R violating interactions re-introduce it, with a nominal value of order the TeV scale.

\item  The full LEFT violates SUSY spontaneously, in a stable minimum, because it does not have the most general terms compatible with R symmetry violation.  From the point of view of LEFT, SUSY breaking ``originates" in the singlet sector, through a variation on the rank mechanism.  In this paper, we have studied only the simplest SUSY breaking model, omitting {\it all} cubic and quadratic terms in the singlet superpotential.  This can be generalized.

\end{itemize}

The details of the above arguments can be found in \cite{tbtjt1} and references cited there.  The present paper has added two important ingredients to our understanding.  If the K\"{a}hler potential of the singlets is canonical, SUSY and electroweak symmetry are still broken, but the SUSY breaking F term belongs to a decoupled chiral superfield, and SUSY breaking is not communicated to the standard model.  The K\"{a}hler potential receives corrections which mix this decoupled singlet with those that do couple to the standard model, from integrating out the two trianons whose masses are above the confinement scale.  In this paper, we performed this integration in the approximation that the masses were much larger than the confinement scale, but ignored the concomitant result that the gauge mediated corrections to chargino masses would be tiny in this limit.  We did this in order to construct a calculable example.  The full K\"{a}hler potential, including strong $SU_P (3)$ gauge corrections to the trianon loops, would have qualitatively similar properties to our unjustified model, but the numerical details would of course be different.

Apart from the qualitative effect of communicating SUSY breaking to the standard model, this approximation enabled us to construct a full sparticle spectrum, which we compared to LHC bounds.  We find that, despite the presence of scalars, which make it easy to get a Higgs boson at $125$ GeV our model required a tuning of parameters of order one part in $20 - 40$, in order to make the electroweak gauge boson masses compatible with the bounds on sparticles.  We note that, while these estimates are purely phenomenological, the underlying logic of the model, following from the gravitino mass formula, puts most of the spectrum in the TeV range.  So the ``tuned" aspect of the model has to do with making the electroweak bosons lighter than the natural scale.  Our numerical approach has not allowed us to guess whether there is a clever solution to this tuning problem, or whether we might have to accept it as a strong coupling accident, like many of the apparent tunings in nuclear physics.

\appendix*
\appendixpage
\section{Discrete R-Symmetry}
In this appendix, we discuss possible R-symmetries that may account for the structure of the above model. The reader should note, that while we have not explicitly discussed inclusions of supergravity in the low-energy effective theory for the Pyramid Scheme, one should consider the effective theory to stem from a microscopic theory of quantum gravity at the Planck scale. As such we do not allow for global continuous R-symmetries, but instead adopt a scheme wherein only discrete R-symmetries are possible. While one may initially view this as an unnecessary  complication, we will show that the system of diophantine equations needed to be solved for a possible R-symmetry is insoluble in the continuum limit and thus only discrete R-symmetries are viable. 

In order to construct an R-symmetry, we must first take into account the allowed operators stemming from the superpotential and then the non-allowed ``dangerous"  baryon and lepton violating higher dimensional operators that we wish to exclude. Additionally, we require that the R-symmetry be anomaly free, which can be guaranteed by requiring that the 'tHooft operators generated by standard model and $SU_P(3)$ instantons vanish. 

With that in mind let us first focus on the terms which must be allowed. These are merely the familiar operators in the pyramid superpotential with added conditions for the inclusion of SM contributions and the neutrino see-saw operator. Altogether the allowed, R-preserving, superpotential operators are  (neglecting couplings):
\begin{equation}
W \supset S_i \mathcal{T}_j \tilde{\mathcal{T}}_j , \, S_i H_u H_d, \, H_u Q \bar{U},  \, H_d Q \bar{D}, \, H_d L \bar{E}, \, (L H_u)^2 \nonumber
\end{equation}
This is equivalent to requiring that the R-charges satisfy the constraints
\begin{eqnarray}
S_i + \mathcal{T}_j + \tilde{\mathcal{T}}_j& = & 2 \nonumber \\
S_i + H_u +H_d &=& 2 \nonumber \\
H_u + Q + \bar{U} &=& 2 \nonumber \\
H_d + Q + \bar{D} &=& 2 \nonumber \\
H_d + L + \bar{E} &=& 2 \nonumber \\
2 ( L + H_u) &=& 2 \nonumber
\end{eqnarray}
where here the fields represent their R-charges and the equations above are to be satisfied modulo $n$. It is already clear that the R-symmetry is unique in the continuum limit, and any additional constraints will require that the R-symmetry be discrete, otherwise the system will have no solution. 

One should notice that we have neglected to include the linear singlet, Higgs mu, quadratic trianon, and cubic trianon operators in the above analysis. This is because these terms explicitly break the discrete R-symmetry and, as discussed previously, are imagined to stem from interactions of the gravitino with the dS horizon \cite{hst-phenom, hst-part-phys}. Such interactions explicitly break any R-symmetry, due to the reflection of the gravitino off the horizon. 

The superpotential and K\"{a}hler terms\cite{pyr-tunnel}, which we want to forbid, either for reasons of minimality or baryon or lepton number violation, are:
\begin{eqnarray}
&&\{ S_i S_j, \, S_i S_j S_k , \,  L L \bar{E}, \, L Q \bar{D}, \, S_i L H_u, \, \bar{U} \bar{D} \bar{D}, \, L H_u, \, Q \bar{U} \bar{E} H_d, \,  Q Q Q L,  \nonumber \\
&& Q Q Q H_d,\, \bar{U} \bar{U} \bar{D} \bar{E}, \, L H_u H_d H_u, \, S_i L L \bar{E}, \, S_i L Q \bar{D}, \, S_i S_j L H_u, \, S_i \bar{U} \bar{D} \bar{D}, \nonumber \\
&& \bar{U} \bar{D}^* \bar{E}, \, H_u^* H_d \bar{E}, \, Q \bar{U} L^*, \, Q Q \bar{D}^* \} \nonumber
\end{eqnarray}
These terms give similar constraints to those above except that we require the sum of the R-charges in the operators to be not equal to 2 (mod $n$) for those terms in the superpotential or 0 (mod $n$) for the K\"{a}hler terms on the bottom row. 

Finally, we have the constraints stemming from instanton generated 'tHooft interactions. The proper conditions such that the R-symmetry is not broken by 'tHooft operators are
\begin{eqnarray}
SU_P(3)^2 U_R(1) &\Longrightarrow&  2 \cdot 3 + 3(\mathcal{T}_1 + \tilde{\mathcal{T}}_1+ \mathcal{T}_2 + \tilde{\mathcal{T}}_2 + \mathcal{T}_3 + \tilde{\mathcal{T}}_3 - 6) = 0 \nonumber \\
SU_C(3)^2 U_R(1) &\Longrightarrow& 2 \cdot 3 + 6(Q-1) +3(\bar{U} + \bar{D} -2) + 3(\mathcal{T}_3 + \tilde{\mathcal{T}}_3 -2) =0 \nonumber \\
SU_L(2)^2 U_R(1) &\Longrightarrow& 2 \cdot 2 + (H_u + H_d -2)+ 9(Q-1) + 3(L-1) + 3( \mathcal{T}_2 + \tilde{\mathcal{T}}_2 - 2) = 0. \nonumber
\end{eqnarray}
Here we will again require that the 'tHooft operator constraints vanish only up to congruency. With all of these constraints in place we can then find the lowest $N$ for which the system of constraints is soluble and solve explicitly for the R-charges in the $\mathbb{Z}_N$ symmetry. 

We find that the lowest order discrete symmetry satisfying all constraints is a $\mathbb{Z}_{12}$ symmetry. The assignment of charges under this group is not unique, but one instance is given by:
\begin{eqnarray}
\bar{E} = &1& \nonumber \\
H_u = H_d = &2& \nonumber \\
\bar{U} = \bar{D} = \mathcal{T}_3= &5& \nonumber \\
T_2 =&6&\nonumber \\
Q = \mathcal{T}_1 =&7& \nonumber \\
\tilde{\mathcal{T}}_1 =&9& \nonumber\\
S_i = \tilde{\mathcal{T}}_2 =&10& \nonumber \\
L = \tilde{\mathcal{T}}_3 = &11& \nonumber
\end{eqnarray}

\bibliographystyle{apsrev4-1}
\bibliography{Pyramid_Scheme_Paper}

\end{document}